\documentclass{JHEP3}

\usepackage{graphicx}
\usepackage{amssymb}
\usepackage[mathscr]{eucal}
\usepackage{amsmath}
\usepackage{cite}
\usepackage{afterpage}

\newcommand{\gsimm}{\raise.3ex\hbox{$>$\kern-.75em\lower1ex\hbox{$\sim$}}}
\newcommand{\lsimm}{\raise.3ex\hbox{$<$\kern-.75em\lower1ex\hbox{$\sim$}}}

\newcommand{\be}{\begin{equation}}
\newcommand{\ee}{\end{equation}}
\newcommand{\ba}{\begin{eqnarray}}
\newcommand{\ea}{\end{eqnarray}}

\newcommand{\bea}{\begin{eqnarray*}}
\newcommand{\eea}{\end{eqnarray*}}

\title{Lectures on Screened Modified Gravity}

\author{Philippe Brax \\
  Institut de Physique Th\'eorique, CEA, IPhT, CNRS, URA 2306,
  F-91191Gif/Yvette Cedex, France \\ E-mail:
  \email{philippe.brax@cea.fr}}

\date{today}

\abstract{The acceleration of the expansion of the Universe has led to the construction of Dark Energy models where a light scalar field may have
a range reaching up to cosmological scales. Screening mechanisms allow these models to evade the tight gravitational tests in the solar system and the laboratory. I will briefly review some of the salient features of screened modified gravity models of the chameleon, dilaton or symmetron types using $f(R)$ gravity as a template\footnote{ Based on lectures given at the Cracow School of Theoretical Physics, Zakopane, Poland, May 2012:
"Astroparticle Physics in the LHC Era".}.}

\begin{document}

\section{Introduction}

The acceleration of the expansion of the Universe\cite{Riess:1998cb,Perlmutter:1998np} has received  no theoretical explanation yet. In these lecture notes, I will present an approach using scalar fields
which can be motivated either from the dark energy\cite{Copeland:2006wr} or the modified gravity points of view\cite{Khoury:2010xi}. I have not attempted to cover the  many aspects of the subject which can be found in very good review papers\cite{Clifton:2011jh}. I have tried to emphasize the unity of  the modified gravity models with a screening mechanism such as chameleons, dilatons and symmetrons, using as a template the example of large curvature $f(R)$ gravity\cite{Hu:2007nk}. I have voluntarily excluded from the analysis the cases of the Galileon\cite{Nicolis:2008in} and massive gravity\cite{deRham:2011rn}, with their associated Vainshtein screening mechanism\cite{Vainshtein:1972sx}. This would have required a special treatment which can be found elsewhere\cite{deRham:2012az}.

\section{Motivation}

The Hubble diagram of supernovae of the type IA first showed that the acceleration parameter
\be
q_0=-\frac{a_0\ddot a_0}{\dot a_0^2},
\ee
where $a_0$ is the scale factor of the Universe\footnote{An index $0$ indicates the present Universe.}, is negative implying that $\ddot a_0>0$ and the Universe accelerates. Assuming that the Cosmological principle stands and that General Relativity (GR) describes the large scale structure of the Universe, the combined data from Cosmic Microwave Background (CMB) observations, large scale structures and Baryon Acoustic Oscillations (BAO) lead to the energy contents of the Universe where $\Omega_{m0} \sim 0.3$ and $\Omega_{\Lambda 0}\sim 0.7$. These two numbers are the energy fraction in matter and dark energy.
Assuming that GR is valid, the Raychaudhury equation yields
\be
\frac{\ddot a}{a}=-\frac{4\pi G_N}{3} ( \rho+ 3 p)
\ee
where $G_N$ is Newton's constant, $\rho$ the energy density of the Universe and $p$ the pressure. Ordinary matter is characterised by an equation of
state
\be
p=w \rho
\ee
 where $w=0$ for non-relativistic matter and $w=1/3$ for radiation. Obviously, ordinary matter with a positive equation of state cannot lead to the acceleration
 of the expansion of the Universe. Hence one must question one of the two hypotheses underlying this result. One may  introduce a new type of energy such that the equation of state becomes negative $w<-1/3$: this new type of matter is called dark energy. On the other hand, if the acceleration of the Universe results from a misunderstanding of the laws of gravity on large scales, gravity must be modified and GR altered.

 In both cases, the acceleration of Universe occurs at very low energy of the order of the critical energy of the Universe
 $\rho_c \sim 10^{-48}\ ({\rm GeV})^4$ implying it should  be describable within the framework of low energy effective field theory. If this were not the case, the
 acceleration of the Universe would be at odds with all the rest of modern particle physics, which underlies the description of the Universe after Big Bang Nucleosynthesis (BBN). Within this context, an effective field theory should be parameterised by a few couplings and masses to be measured experimentally.
 Weinberg's theorem states that the unique low energy field theory of spin 2 particles respecting Lorentz invariance is GR\cite{Weinberg:1965rz}. As Lorentz invariance seems to be a fundamental fact of the Universe, the easiest way of modifying gravity at low energy is to introduce a new field beside the usual graviton. The simplest modification of gravity correspond to adding a mass to the graviton\footnote{The usual Pauli-Fierz theory of massive gravity is ridden with the Boulware-Deser instability in curved space, hence massive gravity is a lot more subtle and recent developments can be found in\cite{deRham:2011rn,deRham:2012az}.}, this automatically generates a field of scalar helicity. Similarly, taking cues from the acceleration of the early Universe, i.e. inflation, dark energy is most easily modeled out using scalar fields. Hence in both the dark energy and the modified gravity cases, scalar fields seem to be compulsory. We will describe how this happens in practice using our first example: $f(R)$ gravity.

\section{ f(R) Gravity}

If GR is not the theory of gravity on large scales, maybe a theory described by
\be
S=\int d^4x \sqrt{-g} \frac{f(R)}{16\pi G_N}
\ee
where $f(R)= R+ h(R)$ (and $h(R)=0$ for GR) is the right one on large scales?
This theory is one amongst many like
\be
S=\int d^4x \sqrt{-g} \frac{f(R,R_{\mu\nu},R_{\mu\nu\rho\sigma})}{16\pi G_N}
\ee
which generically suffer from the Ostrograski instability whereby the Hamiltonian of the model is unbounded from below\cite{Brax:2009ae}.
f(R) gravity does not suffer from this problem. This can be most easily seen by redefining the metric
\be
g_{\mu\nu}= e^{2\beta\phi/m_{\rm Pl}} g^E_{\mu\nu}
\ee
where the metric $g^E_{\mu\nu}$ is the Einstein frame metric.  The Einstein frame is always identified by choosing the  metric allowing one to write the Lagrangian as
\be
S=\int d^4 x \sqrt{-g^E}\frac{R_E}{16\pi G_N}+\dots
\ee
By redefining the metric this way,
as expected, a scalar field appears and the action becomes
\be
S=\int d^4 x \sqrt{-g^E}(\frac{R_E}{16\pi G_N} -\frac{(\partial\phi)^2}{2} -V(\phi))
\ee
where the contractions are to be taken with the Einstein metric.
In this frame, the theory is perfectly well-defined as long as the potential $V(\phi)$ is bounded from below. Its expression is
\be
V(\phi)= \frac{m_{\rm Pl}^2}{2} \frac{ R f_R -f}{f_R^2}
\ee
 and the scalar field is obtained using the mapping
 \be
 f_R\equiv \frac{df}{dR}= e^{-2\beta\phi/m_{\rm Pl}}
 \ee
 where $8\pi G_N=  m_{\rm Pl}^{-2}$ and
 \be
 \beta=\frac{1}{\sqrt 6}.
 \ee
A very useful example, which we will use as a template, consists of the large curvature models\cite{Hu:2007nk} where
\be
f(R)= R-16\pi G_N \rho_\Lambda + \frac{f_{R_0}}{n} \frac{R_0^{n+1}}{R^n}
\ee
for $R\gtrsim R_0$. Expanding to leading order we find that
\be
\frac{\phi}{m_{\rm Pl}}=\frac{f_{R_0}}{2\beta} \frac{R_0^{n+1}}{R^{n+1}}
\ee
and
\be
V(\phi)= \rho_\Lambda(1+4\frac{\beta\phi}{m_{\rm Pl}})-\frac{n+1}{2n} f_{R_{0}} m_{\rm Pl} ^2 R_0(\frac{2\beta \phi}{m_{\rm Pl} f_{R_0}})^{n/(n+1)}+\dots
\ee
Notice that $\rho_\Lambda$ plays the role of a cosmological constant and the potential is a decreasing function of $\phi\lesssim m_{\rm Pl}$ as long as
$R\gtrsim R_0$. In the large curvature regime, this model provides an interesting example of modified gravity. We will investigate  its properties in the course of these lectures.

The acceleration of the Universe occurs when the potential $V(\phi)$ is a dark energy potential.
Generically, dark energy models require that the mass of the scalar field $\phi$ is of the order of the Hubble rate
\be
m_0 \sim H_0 \sim 10^{-43} \ {\rm GeV}.
\ee
This is a very low value implying that the range of the scale field is of the order of the present cosmological horizon. This would not be a problem at all if the scalar field $\phi$ now were decoupled from matter. In most relevant cases, this is not the case  in particular when the $f(R)$ models are defined in the Jordan frame of matter
\be
S=\int d^4x \sqrt{-g} \frac{f(R)}{16\pi G_N} +S_m (\psi, g_{\mu\nu})
\ee
where $S_m$ is the standard model action for matter particles and the coupling of particles to gravity is mediated by the gravitons arising from the metric $g_{\mu\nu}$.
In the Einstein frame, this matter action becomes
\be
S_m (\psi, g_{\mu\nu})\to S_m (\psi, e^{2\beta\phi/m_{\rm Pl}} g_{\mu\nu}^E).
\ee
This implies that fermions couple to the scalar fields as
\be
{\cal L}= \beta \frac{\phi}{m_{\rm Pl}} m_{\psi0} \bar \psi \psi
\ee
where $m_{\psi 0}$ is the mass of the fermion in the Jordan frame.
For a nearly massless scalar fields the propagator  is $-\frac{1}{4\pi} \frac{1}{r}$ and the tree level Feynman diagram
with a scalar exchanged between two fermions gives a potential
\be
\delta\Phi_N=-2 \beta^2 \frac{G_N m_{\psi 0}^2}{r}
\ee
which corrects Newton's potential by a factor $2\beta^2$.
Such a correction is tightly constrained by the Cassini measurement\cite{Bertotti:2003rm}
\be
\beta^2 \le 10^{-5}
\ee
which is violated by $f(R)$ gravity. This does not mean that $f(R)$ gravity is ruled out as the scalar force is screened in dense environments.

\section{Scalar-Tensor theories}

f(R) gravity in the Einstein frame is an example of a scalar-tensor theory defined by
\be
S=\int d^4 x \sqrt{-g^E}(\frac{R_E}{16\pi G_N} -\frac{(\partial\phi)^2}{2} -V(\phi))+S_m (\psi, A^2(\phi) g_{\mu\nu}^E)
\ee
where $A(\phi)$ is an arbitrary function.
The dynamics of these theories are governed by the Einstein equation
\be
R_{\mu\nu}-\frac{1}{2} g_{\mu\nu}R= 8\pi G_N (T^m_{\mu\nu} + T^\phi_{\mu\nu})
\ee
where we have suppressed the $E$ index for convenience. The scalar field has the energy-momentum tensor
\be
T^\phi_{\mu\nu}= \partial_\mu\phi\partial_\nu\phi -g_{\mu\nu}(\frac{(\partial \phi)^2}{2} +V(\phi))
\ee
and the Klein-Gordon equation becomes
\be
D^2 \phi =\frac{\partial V}{\partial \phi} -\frac{\beta_\phi}{m_{\rm Pl}} T^m
\ee
where $T^m$ is the trace of the matter energy momentum tensor.
We have defined the coupling
\be
\beta_\phi= m_{\rm Pl} \frac{\partial \ln A(\phi)}{\partial \phi}.
\ee
Matter is not conserved anymore as
\be
D_\mu T^{\mu\nu}_m= -\frac{\beta_\phi}{m_{\rm Pl}} T^m \partial^\nu \phi.
\ee
This can be better understood in the case of non-relativistic matter
\be
T^{\mu\nu}_m= \rho_E u^\mu u^\nu
\ee
where $u^\mu=\frac{dx^\mu}{d\tau}$ the velocity four-vector, $\tau$ the proper time and $u^\mu u_\mu=-1$.
The non-conservation of matter can be reexpressed as
\be
\dot \rho_E +3 h \rho_E= \frac{\beta_\phi}{m_{\rm Pl}} \rho_E \dot \phi
\ee
where the local Hubble rate is such that $3h=D_\mu u^\mu$, and $\dot \rho_E= u^\mu D_\mu \rho_E$.
This is complemented with the Euler equation
\be
\dot u_\mu= -\frac{\beta_\phi}{m_{\rm Pl}} (\partial_\mu \phi +\dot\phi u_\mu).
\ee
We will come back to this relation later, let us first concentrate on the conservation of matter.
It is particularly useful to define the conserved matter density $\rho$ such
that
\be
\rho_E= A(\phi) \rho
\ee
which satisfies the usual conservation equation for non-relativistic matter
\be
\dot \rho +3 h \rho= 0.
\ee
The Klein-Gordon equation takes then the very convenient form
\be
D^2 \phi =\frac{\partial V_{\rm eff}}{\partial \phi}
\ee
where the effective potential is
\be
V_{\rm eff}(\phi) =V(\phi) +(A(\phi)-1) \rho.
\ee
With a decreasing $V(\phi)$ and an increasing $A(\phi)$, the effective potential acquires a matter dependent minimim
$\phi_{\rm min}(\rho)$ where the mass is also matter dependent $m(\rho)$. As the density increases, the mass increases in such a way that the effects of the scalar field is screened in a dense environment\cite{Khoury:2003aq}. We will come back to this point later.

Let us consider the large curvature $f(R)$ models, in this case we find that the minimum is located at
\be
\phi_{\rm min}(\rho)= \frac{f_{R_0}}{2\beta} m_{\rm Pl} (\frac{\rho_T}{\rho+4\rho_\Lambda})^{n+1}
\ee
where $\rho_T= m_{\rm Pl}^2 R_0= 4\rho_\Lambda + \Omega_{m0}\rho_0$ and $\rho_0=3 H_0^2 m_{\rm Pl}^2$.
Notice that $\phi_{\rm min}(\rho)$ is smaller is denser environments.
When matter is the cosmological matter density which varies as $\rho(a)= \Omega_{m0}\frac{\rho_0}{a^3}$ and the redshift is $1+z=\frac{1}{a}$, the mass at the minimum is simply given by
\be
m(\rho)= m_0 (\frac{4\Omega_{\Lambda 0}+ \Omega_{m0} (1+z)^3}{4\Omega_{\Lambda 0}+ \Omega_{m0}})^{(n+2)/2}
\ee
where
\be
m_0= H_0 \sqrt{\frac{4\Omega_{\Lambda 0}+ \Omega_{m0} }{(n+1) f_{R_0}}}
\ee
Notice that the mass now is larger than the Hubble rate when $f_{R_0}$ is small enough.

\section{Gravitational tests}

Scalar-tensor theories lead to a modification of gravity which can be easily analysed using the Euler equation in the quasi-static limit where $\vert \dot \phi\vert  \ll \vert \partial \phi\vert$. In this case we find that
\be
\dot u_\mu =  -\frac{\beta_\phi}{m_{\rm Pl}} \partial_\mu \phi.
\ee
In the non-relativistic limit where $u^i= v^i$ is the velocity field of the particles, and the metric in the Newton gauge is
\be
ds^2= -(1+2\Phi_N) dt^2 + dx^2 (1- 2\Phi_N)
\ee
where $\Phi_N$ is Newton's potential, we find that the Euler equation reduces to Newton's law
\be
\frac{dv^i}{dt}= -\partial^i \Psi
\ee
where the Newtonian potential is modified
\be
\Psi= \Phi_N + \ln A(\phi).
\ee
Hence gravity is modified by the presence of the scalar field. In the $f(R)$ gravity case we have
\be
\Psi= \Phi_N + \beta \frac{\phi}{m_{\rm Pl}}.
\ee
The same result can be  obtained by writing the relativistic interval in the Jordan frame
\be
ds^2_J= A^2(\phi) ds^2
\ee
as
\be
ds^2= -(1+2\Psi) dt^2 + dx^2 (1- 2\Phi)
\ee
where
\be
\Phi=\Phi_N - \ln A(\phi).
\ee
Hence non-relativistic particles, which couple minimally to the Jordan frame metric feel the presence of two Newtonian potentials $\Phi$ and $\Psi$.
This has important consequences as light bends according to $\Phi+\Psi= 2\Phi_N$ which does not depend on the scalar field while matter particle evolve in the gradient
of $\Psi$. In cosmology, this implies that the velocity field of galaxies is sensitive to $\Psi$ and therefore to the scalar force while cosmic lensing only depends on $\Phi_N$ independently of the scalar field.

Let us now analyse how gravity is modified for spherical and static objects.
\subsection{Point-like particle}

For a point-like particle, the matter density is simply
\be
\rho=m \delta^{(3)}(r)
\ee
where $m$ is its mass. The Klein-Gordon equation for a theory like $f(R)$ gravity where the coupling $\beta$ is constant becomes
\be
\partial^2 \phi= \frac{\beta}{m_{\rm Pl}} m \delta^{(3)}(r)
\ee
implying that
\be
\phi (r)= \frac{\beta}{4\pi m_{\rm Pl} r} m
\ee
and
\be
\Psi= -\frac{m}{8\pi m_{\rm Pl}^2 r} (1+2\beta^2).
\ee
Hence we retrieve the fact that the Newtonian force is larger by a factor of $(1+2\beta^2)$.

\subsection{Small objects}

Let us consider a spherical object of density $\rho_c$ embedded in a sparse environment of density $\rho_\infty$.
Far away from the object, the field converges to the minimum of the effective potential $\phi_\infty$ in the density $\rho_\infty$.
Inside the object, the field is a small perturbation around a value $\phi_0$.
At distances smaller than the range $m_\infty^{-1}$, the field outside reads
\be
\phi= \phi_\infty +\frac{D}{r}
\ee
and the field inside is
\be
\phi= \phi_0 +\frac{V'_0}{m_0^2} (\frac{\sinh m_0r}{m_0 r} -1)
\ee
where $m_0$ is the mass at $\phi_0$ and $V'_0$ the derivative of the effective potential.
The potential outside is defined by
\be
D= \frac{V'_0}{m_0^2}(\cosh m_0 R - \frac{\sinh m_0R}{m_0 R})
\ee
and
\be
\frac{V'_0}{m_0^2} ( 1- \cosh m_0 R)= \phi_\infty -\phi_0.
\ee
When the object is small and its radius $R$ satisfies $m_0 R \ll 1$, we find that
\be
\phi_0 \approx \phi_\infty - 3 \beta m_{\rm Pl} \Phi_N(R)
\ee
where $\Phi_N(R)$ is Newton's potential at the surface of the body.
As the object grows, $\phi_0$ becomes smaller and smaller until it becomes almost uniformly equal to the minimum of the effective potential $\phi_c$ for the inside
density $\rho_c$.
For the small radius case $m_0 R\ll 1$, the total Newtonian potential outside is simply
\be
\Psi= (1+2 \beta^2) \Phi_N(r)
\ee
which is the same result as in the the point-like particle case.

\subsection{Thin shell}

When $m_0 R\gg 1$, the field is almost uniformly equal to $\phi_c$ inside the object, apart from a thin shell at the surface of the object\cite{Khoury:2003aq,Khoury:2003rn}.
The solution outside is
\be
\phi= \phi_\infty -{2 \beta_{\rm eff} m_{\rm Pl}}\Phi_N (R) \frac{R}{r}
\ee
where
\be
\beta_{\rm eff}= 3 \beta \frac{\Delta R}{R}
\ee
and the thin shell extension satisfies
\be
\frac{\Delta R}{R}= \frac{\phi_\infty -\phi_0}{6\beta \Phi_N(R) m_{\rm Pl}}.
\ee
The Newtonian potential outside the object is
\be
\Psi= (1+ 2\beta \beta_{eff}) \Phi_N (r)
\ee
implying that the modification of gravity is screened as long as
\be
\vert{\phi_\infty -\phi_0}\vert \le {2\beta \Phi_N(R) m_{\rm Pl}}.
\ee
This is the screening criterion for models like $f(R)$ gravity with a constant $\beta$. In fact, the same criterion holds
for all models with screening such as chameleons, dilatons and symmetrons where $\beta$ should be replaced by $\beta_\infty$\cite{Brax:2012gr}.

In practice, when one applies these results to the Cassini experiment, one must impose that
\be
\beta_\infty \beta_{\rm eff} \le 10^{-5}
\ee
Another type of constraints can be obtained from cavity tests of the gravitational force.

Inside a cavity of radius $R$ filled with a vacuum density $\rho_\infty$ and surrounded by a bore of density $\rho_c$, the field is essentially constant close to the
centre of the cavity with a value $\phi_0$ determined by
\be
1+ \frac{\sinh m_0 R}{m_0R}= -\frac{\phi_0 m_0^2}{V'_0}.
\ee
When this condition is satisfied, the test bodies inside the cavity must be screened and the effective coupling satisfy
\be
\beta_{\rm eff}^2\le 10^{-5}
\ee
where
\be
\beta_{\rm eff}\approx \frac{\phi_0}{2 \Phi_N(R)  m_{\rm Pl}}
\ee
and $\Phi_N$ is the Newtonian potential of the test particles.

In the case of large curvature $f(R)$ gravity, the Cassini constraint reads
\be
f_{R_0}\le 10^{-5} \Phi_N({\rm sun})(\frac{\rho_\infty}{\rho_T})^{n+1}
\ee
where $\rho_\infty \sim 10^6 \rho_T$ and  $\Phi_N({\rm sun})\sim 10^{-6}$ implying that
\be
f_{R_0}\le 10^{6n-5}.
\ee
This is a very mild constraint. The cavity constraint is even milder as there is no solution for $\phi_0$ implying that $\phi_c$ is the value inside the cavity and
therefore the test particles are effectively decoupled from the scalar field.

\section{Models}

Large curvature $f(R)$ gravity is not the only type of models of interest. We will sketch three different ones here.
\subsection{Chameleons}

Chameleons\cite{Khoury:2003aq,Khoury:2003rn,Brax:2004qh} share the same coupling as $f(R)$ models with
\be
A(\phi)= e^{\beta\frac{\phi}{m_{\rm Pl}}}
\ee
where $\beta$ is a free parameter. Typically, the potential for chameleon models can be taken of the inverse power law form with
\be
V(\phi)= \Lambda^4 + \frac{\Lambda^{n+4}}{\phi^n}
\ee
where $\Lambda \sim 10^{-3}$ eV both acts like a cosmological constant and implies that the Cassini and cavity bounds are satisfied thanks to the thin shell effect.

\subsection{Dilatons}

Gravity tests are evaded in a very different way in the dilatonic models\cite{Brax:2010gi}. This is due to the shape of the coupling function
\be
A(\phi)=1+ \frac{A_2}{2m_{\rm Pl}^2}(\phi-\phi_\star)^2
\ee
which implies that
\be
\beta_\phi= \frac{A_2}{m_{\rm Pl}} (\phi-\phi_\star)
\ee
and the coupling to matter converges to zero when $\phi$ is stabilised close to $\phi_\star$ in a dense environment.
The potential for dilaton models must be smooth, positive and slowly varying like
\be
V(\phi)=V_0 e^{-\phi/m_{\rm Pl}}
\ee
where $V_0$ gives the order of magnitude of the energy density required to generate the acceleration of the Universe.
The minimum of the effective potential $\phi_{\rm min}(\rho)\to \phi_\star$ in a dense environment where $\beta (\phi_\star)=0$.

\subsection{Symmetrons}
The symmetrons\cite{Hinterbichler:2010es,Olive:2007aj,Pietroni:2005pv} have the same type of coupling as the dilatons where
\be
A(\phi)=1+ \frac{A_2}{2m_{\rm Pl}^2} \phi^2
\ee
implying that
\be
\beta_\phi= \frac{A_2}{m_{\rm Pl}} \phi.
\ee
The potential is a Mexican hat with
\be
V(\phi)= V_0 - \frac{\mu^2}{2} \phi^2 +\frac{\lambda}{4} \phi^4
\ee
implying that there is a symmetry breaking transition for a density
\be
\rho_\star= \frac{\mu^2 m_{\rm Pl}^2}{A_2}
\ee
for which in an environment with  $\rho\ge \rho_\star$, the minimum is at the origin with a vanishing coupling
\be
\phi_{\rm min}(\rho)=0\to \beta_\phi(\phi_{\rm min})=0.
\ee
Hence gravity is only modified at low density, corresponding to the late time Universe or sparse environments.

\section{ Cosmology}

The cosmology of screened models of modified gravity is universal at the background level and differs only for perturbations.
\subsection{Background Cosmology}

When the mass of the scalar field at the minimum of the effective potential is large enough, i.e. such that
$m(\rho)\gg H$, the minimum is stable. This implies that if the field settles at the minimum early enough in the
Universe, it will stay there for the rest of the evolution of the Universe.
This is particularly important as the particle masses have a $\phi$ dependence
\be
m_\psi= A(\phi) m_{0\psi}
\ee
where $m_{0\psi}$ is the mass in the Jordan frame. If $A(\phi)$ varied abruptly during Big Bang Nucleosynthesis (BBN), the formation of the elements
would be altered. If the field follows the time evolution of the minimum before BBN, its stability is guaranteed especially when species like the electron decouple leading to a jump of the trace of the energy momentum tensor. If the field were not at the minimum, this would lead to kicks to the field values and therefore a large variation
of the particle masses. This is not the case when the field follows the minimum from a redshift around $z\sim 10^{10}$.

At the background level, the energy density of the scalar field is
\be
\rho_\phi= \frac{\dot \phi^2}{2} + V_{\rm eff}(\phi)
\ee
 and the pressure
\be
p_\phi= \frac{\dot \phi^2}{2}- V(\phi).
\ee
The effective equation of state\cite{Brax:2012gr}
\be
1+w_\phi= 1+ \frac{p_\phi}{\rho_\phi}\approx \frac{\dot\phi^2}{V} +(A(\phi)-1)\frac{\Omega_m}{\Omega_\phi}
\ee
is extremely close to $-1$ as
\be
\frac{\dot\phi^2}{V}= 27 \Omega_m \beta_\phi^2 (\frac{H^4}{m(\rho)^4})\frac{\Omega_m}{\Omega_\phi}
\ee
 and
\be
(A(\phi)-1)\frac{\Omega_m}{\Omega_\phi}\approx 3\Omega_m \beta_\phi^2 (\frac{H^2}{m(\rho)^2})\frac{\Omega_m}{\Omega_\phi}
\ee
implying that
\be
1+w_\phi={\cal O}(\frac{H^2}{m(\rho)^2})
\ee
which is tiny number, and therefore these models behave like $\Lambda$-CDM as long a $\Omega_\phi$ is not too small. This is in particular the case in the late time Universe.

\subsection{Perturbations}

The only hope of distinguishing the screened modified gravity models from $\Lambda$-CDM resides in the very different properties of cosmological perturbations.
Let us come back to the conservation equation which we will study in the Newtonian gauge
\be
ds^2= a^2(\eta)(-(1+2\Phi_N) d\eta^2 + dx^2 (1- 2\Phi_N)).
\ee
Putting
\be
u^0=a^{-1}(1-\Phi_N),\ u^i=a^{-1} v^i
\ee
and
\be
h= \frac{ {\cal H}}{3}+ \frac{ \theta}{3a}
\ee
where $\theta=\partial_i v^i$ and ${\cal H}=a'/a$ with $'=d/d\eta$, we find that
\be
\delta'=-\theta
\ee
with $\delta= \frac{\delta\rho}{\rho}$.

Similarly the Euler equation becomes
\be
v_i'+{\cal H} v_i + v^j\partial_j v^i =-\partial \Phi_N -\frac{\beta_\phi}{m_{\rm Pl}} \partial_i \phi
\ee
which we  linearise to obtain
\be
\theta' +{\cal H}\theta = k^2 \Phi_N +\frac{\beta_\phi}{m_{\rm Pl}} k^2 \delta\phi
\ee
in Fourier modes.
Neglecting the energy density of the scalar field, the Poisson equation becomes
\be
k^2 \Phi_N= - \rho \frac{a^2 \delta}{2m_{\rm Pl}^2}.
\ee
In the quasi-static approximation, the Klein-Gordon equation becomes an algebraic relation
\be
\frac{\delta\phi}{m_{\rm Pl}}= -\frac{\beta_\phi \rho \delta}{\frac{k^2}{a^2} + m^2(\rho)}.
\ee
Combining these equations we find that the density contrast must satisfy\cite{Brax:2004qh}
\be
\delta'' +{\cal H} \delta' -\frac{3\Omega_m{\cal H}^2}{2} (1+ \epsilon(k,a))\delta=0
\ee
where
\be
\epsilon(k,a)= \frac{ 2\beta_\phi^2}{ 1+ \frac{m^2(\rho)a^2}{k^2}}
\ee
captures the effects of modified gravity at the linear level. It can be immediately inferred that on large scales $k\ll {a m(\rho)}$, the evolution of the density contrast is the same as in the $\Lambda$-CDM paradigm, i.e. in the matter dominated era it follows
\be
\delta \sim a(\eta).
\ee
In the radiation era, the density contrast grows logarithmically still. The main difference occurs on small scales inside the Compton wavelength of the scalar field
$k\gg am(\rho)$. In this regime the growth is enhanced with a growing modes scaling like
\be
\delta \sim a^{\nu/2}
\ee
where
\be
\nu= \frac{-1+\sqrt{1+24(1+\beta_\phi^2)}}{2}.
\ee
This results is of course only valid when $\beta_\phi$ is constant, on the other hand the anomalous growth is present for all models even when
$\beta_\phi$ depends on the scale factor via the evolution of the field $\phi_{\rm min}(\rho)$.
We will see in the next section that the mass scales are  such that the main effects of modified gravity arise below 10 Mpc where non-linear effects must be taken into account. This is investigated via N-body simulations, showing a distorsion of the matter power spectrum at the percent level which may be within reach with forthcoming
surveys such as EUCLID.

\section{Modified Gravity Tomography}

The screened models that we have presented so far are all defined by a Lagrangian and two functions $V(\phi)$ and $A(\phi)$. The path from this Lagrangian formulation to observations such as gravity tests or the growth of large scale structure is not direct. From a phenomenological point of view, it would be more efficient  to define models using quantities which are closer to the physical observables. This is easily realised if the mass $m(a)$ and the coupling constant $\beta(a)$ are given as a function of the scale factor of the Universe. Indeed these two functions completely characterise the time evolution of the linear
cosmological perturbations. In particular, the mass $m(a)$ indicates which scales are affected by the anomalous growth of structure due to the scalar field.
In terms of matter density, the cosmological matter density $\rho(a)$ varies from $10\ {\rm g/cm^3}$ at redshifts around $z=10^{10}$ before BBN to cosmological densities now for $z=0$. Hence this parameterisation is in one to one correspondence with all the range of densities accessible to experiments in the solar system and in astrophysics \footnote{Larger densities like in neutron stars require to know the mass and the coupling for a much large $z$, this is not necessary to what follows.}.
In fact the knowledge of $m(a)$ and $\beta(a)$ is entirely enough to fully define the non linear functions $V(\phi)$ and $\beta_\phi$. This is simply a reconstruction mapping which is a mathematical property, it has nothing to do with the fact that the field follows or not the minimum of the effective potential since before BBN.

The field values can be parametrically defined by the following expression\cite{Brax:2011aw}
\begin{equation}
\phi(a)=  \frac{3}{m_{\rm Pl}}\int_{a_{\rm ini}}^a \frac{\beta (a)}{a m^2(a)}\rho (a)  da +\phi_c,\label{phi}
\end{equation}
where $\phi_c$ is the value of the field at the minimum in a dense region of density $\rho(a_{\rm ini})$. Taking $a_{\rm ini} \sim 10^{-10}$ implies that this value
of the field is the one inside dense bodies on earth where $\rho(a_{\rm ini})\sim\ 10\ {\rm g/cm^3}$. The value of $\phi(a)$ is the minimum one in a body of density
$\rho (a)$.
Similarly the potential can be obtained using
\begin{equation}
V(a)=V_0 -3  \int_{a_{\rm ini}}^a \frac{\beta(a)^2}{am^2(a)} \frac{\rho^2}{m^2_{\rm Pl}} da.
\label{V}
\end{equation}
where $V_0$ is a constant. Eliminating $a$ from these two expressions, one obtains the potential $V(\phi)$ and the coupling  $\beta_\phi$.
This method is particularly important for numerical simulations as models defined by $m(a)$ and $\beta (a)$ can be easily analysed in their non-linear regime
using the reconstruction mapping.

This method is also very efficient to impose the screening condition. Indeed it can be reformulated as
\be
\vert \phi_G -\phi_0\vert \lesssim 2 \beta_0 \Phi_G
\ee
where $\phi_0$ is the value of the minimum far away from the object where the coupling is $\beta_0$ and the object has a Newtonian potential at its surface
equal to $\Phi_G$ and a value at the minimum  equal to $\phi_G$. This can be reexpressed a
\be
 \frac{9}{2}\Omega_{m0}\frac{H_0^2}{m^2_0} \int_{ a_{G}}^{a_0} da \frac{g(a)}{a^4 f^2(a)}\le \Phi_G
\ee
where we have defined $m(a)=m_0 f(a)$ and $\beta(a)= \beta_0 g(a)$.
Let us apply this inequality to the Milky Way which must be screened to avoid a disruption of the dynamics of the galactic halo. In this case
$a_G=10^{-2}$ corresponding to a density inside the galaxy which is $10^6$ larger than the cosmological density and $a_0=1$ assuming that the Milky Way is surrounded by the cosmological vacuum. For models like $f(R)$ gravity in the large curvature regime where the integral $\int_{ a_{G}}^{a_0} da \frac{g(a)}{a^4 f^2(a)}={\cal O}(1)$,
and upon using that the galactic Newtonian potential is $\Phi_G\sim 10^{-6}$, we get that\cite{Brax:2011aw,Wang:2012kj}
 \be\frac{m_0}{H_0} \gtrsim 10^{3}.\ee
This condition is independent of $\beta_0$ and means that any screened modified gravity model will have effects on Mpc scales only.
For the case of large curvature $f(R)$ gravity, this implies that\cite{Hu:2007nk}
\be
f_{R_0}\le 10^{-6}
\ee
This is a loose bound as the screening of the Milky Way is not a strong quantitative constraint. A slightly stronger bound $f_{R_0}\le 5\cdot 10^{-7}$ comes from distance indicators of screened vs unscreened astrophysical objects \cite{Jain:2012tn}.

\section{Quantum Corrections}

\subsection{Effective field theory}

The modified gravity models with a screening property are at best effective field theories valid below the electron mass where all the massive fields of the standard model of particle physics have been integrated out\footnote{Apart from the neutrinos.}. This is an appropriate description of cosmology after BBN as the matter particles can be appropriately modeled using a fluid approximation. As a result the only quantum fluctuations which are present in this low energy model are the ones of the scalar field itself whose mass is extremely low below the cut off around the MeV scale.
To evaluate these quantum corrections one must expand the field around the background value $\phi_0$ of the scalar field in the presence of matter. The effective potential reads then
\be
V_{\rm eff}(\phi_0+\delta \phi)= V_0 + \frac{m_0^2}{2}\delta\phi^2 +  \phi_0^4 \sum_{p\ge 3} c_p (\frac{\delta \phi}{\phi_0})^p
\ee
where $c_p$ are dimensionless coupling constants
\be
c_p= \phi_0^{p-4}\frac{V^{(p)}(\phi_0)}{p!}.
\ee
This expansion can also be written
\be
V_{\rm eff}(\phi_0+\delta \phi)= V_0 + \frac{m_0^2}{2}\delta\phi^2 + \Lambda_3 \delta\phi^3 + \frac{\lambda}{4!} \delta\phi^4+ \sum_{p\ge 5} \frac{\delta\phi^p}{\Lambda_p^{p-4}}
\ee
where each of the scales $\Lambda_p,\ p>4$ act as a cut-off scale for each of the higher order operators. For the non-renormalisable operators $p>4$, these scales
give an effective order of magnitude of the effective cut off of the theory as processes at energies larger than $\Lambda_p$ would violate the unitarity of the theory. The validity of 
perturbation theory also requires that $\vert \lambda \vert \le 1$.

Let us consider a simple example where
\be
V(\phi)= V_0 +\frac{\beta V_1}{m_{\rm Pl}} \phi+ \epsilon \Lambda_0^4 (\frac{\phi}{\phi_T})^\alpha
\ee
where $\epsilon=-1$ when $\alpha>0$ and vice versa.
In the large curvature $f(R)$  gravity case, we have the identification
\be
V_0= \rho_\Lambda,\ V_1=4\rho_\Lambda,\  \alpha=\frac{n}{n+1},\ \Lambda_0^4=\frac{n+1}{2n} f_{R_{0}} m_{\rm Pl} ^2 R_0,\ \phi_T=\frac{f_{R_0} m_{\rm Pl}}{2\beta}.
\ee
For a constant coupling $\beta$, we find that the minimum of the effective potential is located at
\be
\phi_0=\phi_T(\frac{\beta (\rho+V_1) \phi_T}{\alpha m_{\rm Pl} \Lambda_0^4})^{1/(\alpha -1)}.
\ee
For these models we have
\be
\Lambda_p^{p-4}= \frac{p!}{4! \epsilon (\alpha-4)\dots (\alpha-p+1)}\frac{\phi_0^{p-4}}{\lambda}
\ee
and
\be
\vert \lambda\vert =(\alpha-2)(\alpha-3)\frac{m_0^2}{\phi_0^2}
\ee
where
\be
m_0^2=\epsilon \alpha (\alpha -1) \frac{\Lambda_0^4}{\phi_T^2} (\frac{\phi_0}{\phi_T})^{\alpha-2}.
\ee
As a result, the validity of
perturbation theory is guaranteed when
\be
m_0\lesssim \phi_0
\ee
and the effective cut-off is
\be
\Lambda_p \sim \phi_0.
\ee
Therefore he validity of
perturbation theory is not violated  as long as the scalar field is light enough compared to the effective cut off. This guarantees the consistency
of the model as the scalar field does not need to be integrated out to obtain the low energy effective theory below the cut off scale $\phi_0$.
We have also
\be
\frac{m_0^2}{\phi_0^2}= \vert (\alpha-1)\vert \frac{\beta(\rho+V_1)}{\phi_T \phi_0^3}
\ee
and one can directly see that the theory is strongly coupled   at high enough density.

\subsection{Coleman-Weinberg corrections}

The one loop correction to the scalar potential at the one loop level is given by
\be
\delta V= \frac{m_0^4}{64\pi^2} \ln \frac{m_0^2}{\mu^2}
\ee
where $\mu$ is a renormalisation scale which can be taken at the dark energy scale $\mu\sim 10^{-3}$ eV.
This new potential may upset the properties of the screened modified gravity models\cite{Upadhye:2012vh} if its order of magnitude is larger than the dark energy scale
now, if the minimum of the potential is shifted or even disappears and finally if the mass at the new minimum is much larger than $m_0^2$.
The first problem is easily solved as long as $m_0/H_0$ is not large, say $10^3$, as the Hubble rate now is so small.
The order of magnitude of the correction to the first derivative of the potential at $\phi_0$ is given by
\be
\delta V'_0 \sim \frac{m_0^4}{\phi_0}\sim \lambda V'_0.
\ee
As long as perturbative unitarity with $\vert \lambda\vert  \ll 1$ is preserved, the shift of the slope of the potential is negligible.
Similarly we have
\be
\delta m_0^2 \sim \lambda m_0^2
\ee
and the mass shift is also tiny when perturbative unitarity is valid.
Hence we find that as long as $\vert \lambda \vert \ll 1$, which is necessary to guarantee that the effective field theory makes any sense, the one loop corrections due to the scalar field are irrelevant.

There is only one type of fermion which has not been integrated out, i.e. the  neutrinos. The masses of neutrinos depend on the scalar field according to
\be
m_\psi= A(\phi)m_{\psi 0}
\ee
implying that the one loop contribution to the scalar potential is
\be
\delta V\approx -\frac{(m_{\psi 0})^4}{64\pi^2}(1+4\frac{\beta\phi}{m_{\rm Pl}}).
\ee
This leads to a renormalisation of
\be
V_{0,1}\to V_{0,1}-\frac{(m_{\psi 0})^4}{64\pi^2}
\ee
for each neutrino species. As $m_{\psi 0}\lesssim 10^{-3}$ eV in most probable scenarios of neutrino physics, we find that the effect of the neutrinos
is to renormalise the cosmological constant by a value compatible with its observable value.

\section{Lorentz Violation}

We have focused so far on modified gravity where a scalar field couples to matter in a "conformal" way via the coupling function $A(\phi)$.
This is  not the only possibility by far and another type of coupling includes a disformal term\cite{Wyman:2011mp,Brax:2012hm}. It involves the matter action
$
S_m(\psi, g_{\mu\nu})
$
where matter couples to the  metric
\begin{equation}
g_{\mu\nu}=A^2(\phi)(g_{\mu\nu}^E+\frac{2\partial_\mu \phi\partial_\nu\phi}{M^4}),
\end{equation}
$M$ is a suppression scale and $g_{\mu\nu}^E$ the Einstein metric with which the Einstein-Hilbert term is written.

Defining the energy momentum tensor  as
$
T^{\mu\nu}=\frac{2}{\sqrt{-g}} \frac{\delta S_m^i}{\delta g_{\mu\nu}}
$
and expanding the action to linear order, we find that the scalar field couples derivatively to matter
\begin{eqnarray}
&&S = \int d^4x\sqrt{-g_E}\left\{\frac{m_{\rm Pl}^2}{2}{
R_E}-\frac{(\partial \phi)^2}{2}- V(\phi) \right\}\nonumber\\
&& +\int d^4x\sqrt{-g_E} \frac{\partial_\mu \phi\partial_\nu \phi}{ M^4} T^{\mu\nu}+  \int d^4x \sqrt{- g}A^4(\phi) {\cal
L}_m(\psi,A^2 (\phi) g_{\mu\nu}^E)\,, \label{action}
\end{eqnarray}
As soon as $\theta_{\mu\nu}=\partial_\mu\phi\partial_\nu\phi$ does not vanish due to the presence of matter,  Lorentz invariance is broken and
a  Lorentz violating coupling to the matter energy momentum tensor is present in the model.
In a static configuration and in the presence of non-relativistic matter, the disformal coupling has no effect and therefore cannot be constrained by
static tests of modified gravity. On the other hand, it can modify the propagation of fermions in dense environments.

The action for
massless fermions is
\be
S_F=- \int d^4x \sqrt{-g} \frac{i}{2}(\bar \psi \gamma^\mu D_\mu \psi - (D_\mu \bar \psi) \gamma^\mu \psi)
\ee
leading to  the energy momentum tensor
\be
T^F_{\mu\nu}= -\frac{i}{2} (\bar \psi \gamma_{(\mu} D_{\nu)} \psi - (D_{(\mu }\bar \psi \gamma_{\nu)} \psi)
\ee
symmetrised over the indices. This induces the following interaction terms with the scalar field
\begin{equation}
-\frac{i}{2}\int d^4 x \sqrt{-g_E} \frac{\partial^\mu \phi \partial^\nu\phi}{M^4} (\bar \psi \gamma_{(\mu} D_{\nu)} \psi - (D_{(\mu }\bar \psi \gamma_{\nu)} \psi).
\end{equation}
In a typical situation where the metric is Minkowskian to a good approximation and  the scalar field is static, the interaction term reduces to
\begin{equation}
-\frac{i}{2}\int d^4 x  \  d^i d^j (\bar \psi \gamma_{i} \partial_j \psi- \partial_i \bar \psi \gamma_j \psi)
\end{equation}
where
$
d^i= \frac{\partial^i\phi}{M^2}
$
is a slowly varying function of space only.
Hence a static configuration of the scalar field yields a Lorentz violating interaction in the Fermion Lagrangian.
The resulting Dirac equation becomes
\begin{equation}
i(-\gamma^0\partial_0 +\gamma^i \partial_i +d^i d^j \gamma_i \partial_j)\psi=0.
\end{equation}
The dispersion relation is obtained by squaring the modified Dirac operator to obtain
\begin{equation}
p_0^2 = (c^2)^{ia}p_i p_a.
\end{equation}
This becomes the dispersion relation in an anisotropic medium with a square velocity tensor
\begin{equation}
(c^2)^{ia}=(\delta^{ij}+ d^i d^j)(\delta^{aj} + d^a d^j).
\end{equation}
The eigenmodes of the velocity tensor are $d^i$ and two vectors $e^i_\lambda, \lambda=1,2$ orthogonal to $d^i$.
The eigenspeeds are $c_d=(1+\vert d\vert^2)$ and twice $c_\lambda=1$. Hence, fermions go  faster than light in the direction of the
gradient $\partial^i \phi$, i.e. along the scalar lines of force,  with
\begin{equation}
\Delta c\equiv c_d-1= \vert d\vert^2.
\end{equation}
This is of course something which has never been observed! This implies that the scale $M$ must be large enough as the velocity of fermions could be larger than the speed of light in the thin shell of a screened object. This is very reminiscent of the wrongly announced result by the OPERA collaboration\dots

\section{Conclusion}

The models of screened modified gravity evade all gravitational tests. One may wonder how they could be efficiently probed. One possibility would be
to find relevant situations where screening is not efficient. On astrophysical scales, this may happen in the Mpc range and may lead to a bump
in the deviation of the power spectrum of density fluctuations from the $\Lambda$-CDM one\cite{Oyaizu:2008tb,Brax:2012nk}. The physics of stars, screened vs unscreened is also interesting\cite{Davis:2011qf}. Satellite tests of the equivalence principle may also be sensitive to an unscreened force\cite{Khoury:2003aq}. In the laboratory, the Casimir effect experiments are extremely sensitive probes of new forces\cite{Brax:2010xx}. Finally, the interaction of slow neutral particles with matter, i.e. neutrons which are generically unscreened, may reveal surprises in low energy particle physics\cite{Brax:2011hb}. All in all, the search for modified gravity effects opens up a new era for physics at the low energy frontier.
\section{Acknowledgments}
I would like to thank the organisers of the Zakopane summer school for their invitation to give these lectures. J. Sakstein and H. Winther have been kind enough to comment on the manuscript. I am also extremely grateful to all my collaborators over all these years. I finally apologise for a very incomplete list of references.

\bibliography{zako}

\end{document}